# 8

# Probing Brain Oxygenation Waveforms with Near Infrared Spectroscopy (NIRS)

Alexander Gersten[1], Jacqueline Perle[2],
Dov Heimer[3], Amir Raz[4] and Robert Fried[5]
[1]*Department of Physics, Ben-Gurion University, Beer-Sheva*
[2]*Department of Psychology,
Hunter College of the City University of New York, NY*
[3]*Pulmonary Unit, Soroka University Hospital, Beer-Sheva*
[4]*McConnell Brain Imaging Centre, Montreal Neurological Institute,
McGill University, Montreal*
[5]*City University of New York, NY*
[1,3]*Israel*
[2,5]*USA*
[4]*Canada*

## 1. Introduction

Breathing may have dramatic effects on the human cerebral blood flow (CBF) and cognition. This was already known long time ago to Chinese (Li Xiuling, 2003), Hindus (Kuvalayananda, 1983) and Tibetans (Mullin, 1996). The main parameter influencing the CBF is the arterial partial pressure of carbon dioxide ($PaCO_2$). About 70% increase in $PaCO_2$ may double the blood flow (Sokoloff, 1989, Guyton, 1991). The increase of blood flow to the brain results in an increase of nutrients and oxygen, which may influence to great extent brain's physiology.

On the other hand lowering the $PaCO_2$ through hyperventilation found an application in neurosurgery (Feihl and Perret, 1994).

An evaluation of the CBF dependence on $PaCO_2$ was given by Gersten (Gersten, 2011). The human normal value of $PaCO_2$ is about 40 mmHg. Fig. 1 depicts our estimation of the changes of CBF as a result of changing $PaCO_2$ from normal. The estimate is based on the data of Refs. (Reivich 1964, Ketty and Schmidt 1948, Raichle et all 1970). An increase of only 12.5% from normal in $PaCO_2$ leads to the state of hypercapnia ($PaCO_2$ above 45 mm Hg).

With aging both CBF and $PaCO_2$ tends to fall down, therefore breathing exercises (or procedures) with the aim to increase $PaCO_2$ and CBF might be important in preventing and alleviating neurodegenerative diseases. They may be also important in improving intellectual abilities. The amount of $PaCO_2$ in our arteries is proportional to the $CO_2$ which was metabolized by the organism and inversely proportional to the lung ventilation (i.e. to the amount of air exchanged between the lungs and the environment in one minute).

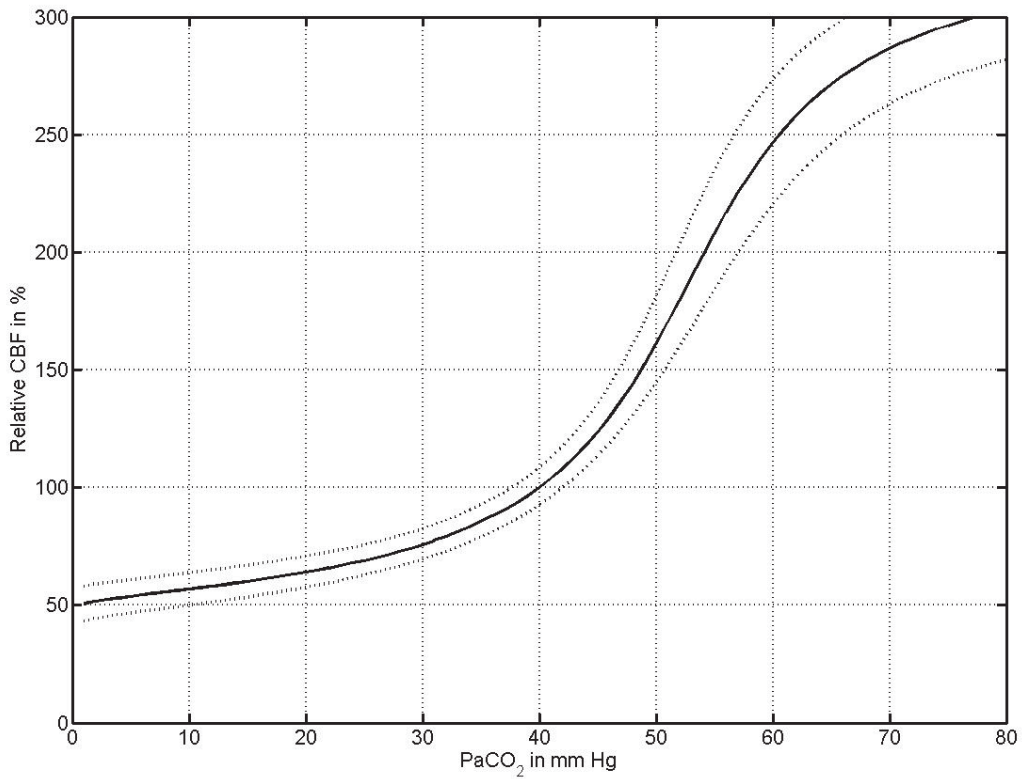

Fig. 1. Estimated changes of human CBF from normal (100%) against $PaCO_2$. Normal $PaCO_2$ =40 mm Hg is assumed, it corresponds to ordinate value of 100%. From (Gersten, 2011).

An instantaneous way to change $PaCO_2$ can be accomplished by influencing the ventilation. Respiratory periodicities were observed in fMRI, but they were treated as artifacts (Windischberger et al., 2002) or as a noise (Raj et al., 2001). The main research effort was directed towards the elimination of the effect of the respiratory periodicities. Our aim is to study the effect of respiration on brain's oxygenation and performance.

Near infrared spectroscopy (NIRS) is an efficient way to probe brain's oxygenation (Rolfe, 2000; Thavasothy et al., 2002; Elwell et al., 1996). Brain oxygenation increases with an increase of $PaCO_2$ (Imray et al.,2000, Thavasothy et al., 2002). In this chapter we examine the dependence of brain's oxygenation on $PaCO_2$.

The technique of near infrared spectroscopy (NIRS) allows to measure the oxygenation of the brain tissue. The particular problems involved in detecting regional brain oxygenation (rSO2) are discussed. The dominant chromophore (light absorber) in tissue is water. Only in the NIR light region of 650-1000 nm, the overall absorption is sufficiently low, and the NIR light can be detected across a thick layer of tissues, among them the skin, the scull and the brain. In this region, there are many absorbing light chromophores, but only three are important as far as the oxygenation is concerned. They are the hemoglobin (HbO2), the deoxy-hemoglobin (Hb) and cytochrome oxidase (CtOx). In the last 20 years there was an enormous growth in the instrumentation and applications of NIRS.

The devices that were used in our experiments were : Somanetics's INVOS Brain Oximeter (IBO) and Toomim's HEG spectrophotometer. The performances of both devices were compared including their merits and drawbacks. The IBO is based on extensive efforts of an R&D group to develop a reliable device, which measures well the rSO2. It is now used efficiently in operating rooms, saving human lives and expenses. Its use for research however has two drawbacks: the sampling rate is too small and the readings are limited to only two significant digits. The HEG device does not have these drawbacks, but is not developed sufficiently at this time to measure rSO2. We have measured the HEG readings and compared them with the rSO2 readings of the IBO. Our findings show that the HEG can be used to measure relative changes of rSO2.

Results of an experiment are presented whose aim is to explore the relationship between respiration and cerebral oxygenation. Measurements of end tidal $CO_2$ ($EtCO_2$) were taken simultaneously with cerebral oxygen saturation ($rSO_2$) using the INVOS Cerebral Oximeter of Somanetics. Due to the device limitations we could explore only subjects who could perform with a breathing rate of around 2/min or less. Six subjects were used who were experienced in yoga breathing techniques. They performed an identical periodic breathing exercise including periodicity of about 2/min. The results of all six subjects clearly show a periodic change of cerebral oxygenation with the same period as the breathing exercises. Similar periodic changes in blood volume index were observed as well.

We tested the hypothesis that simple breathing exercises may significantly increase cerebral blood flow (CBF) and/or cerebral oxygenation. Eighteen subjects ranging in age from nineteen to thirty nine participated in a four-stage study during which measurements of end tidal $CO_2$ (EtCO2 - by capnometer) and local brain oxygenation (by near-infrared spectroscopy (NIRS) sensor) were taken. The four stages were 1) baseline, 2) breathing exercises, 3) solving an arithmetic problem, and 4) biofeedback. During the breathing exercises there was a significant increase in $EtCO_2$ indicating a significant increase in global CBF. The increase in global CBF was estimated on the basis of a theoretical model. During the arithmetic and biofeedback tasks there was a significant increase in the local (Fp1) oxygenation, but it varied between the different participants. The results may lead to new clinical applications of CBF and brain oxygenation monitoring and behavioral control.

## 2. Fundamentals of NIRS

The study of the human brain made a big step forward with the introduction of noninvasive techniques, among them the near-infrared spectroscopy (NIRS). This technique allows to measure the oxygenation of the brain tissue (Alfano et al., 1997, 1998; Chance 1998, 1998a; Delpy and Cope, 1997; Hoshi, 2003; Obrig, 2003; Rolfe, 2000; Strangman et al., 2002).

The light, with wavelengths 650-1000 nm, penetrates superficial layers of the human body, among them the skin, the scull and the brain. It is either scattered within the tissue or absorbed by absorbers present in the tissue (chromophores).

The visible light has wavelengths 400-700 nm, some individuals can see up to 760 nm. Formally, the red light extends within the wavelengths of 630-760 nm, and the near infrared light within 760-1400 nm. However, these terms are not precise, and are used differently in various studies. Here we will refer to the near infrared light as having the spectrum of 650-1000 nm, i.e. the light that penetrates superficial layers of the human body.

The dominant chromophore in tissue is water. It absorbs strongly below 300 nm and above 1000 nm. The visible part of the light spectrum, between 400 and 650 nm, is almost non-transparent due to strong absorption of hemoglobin and melanin. Only in the NIR light region of 650-1000 nm, the overall absorption is sufficiently low, and the NIR light can be detected across a thick layer of tissue.

In the following pages, we will be concerned with utilizing the NIR light to determine the oxygenation of the brain tissue. In the rather transparent NIR region, there are many absorbing light chromophores, but only three are important as far as the oxygenation is concerned. They are the hemoglobin ($HbO_2$), the deoxyhemoglobin (Hb) and cytochrome oxidase (CtOx), Hb and $HbO_2$ (which carries the oxygen) are found inside the red blood cells. CtOx is the enzyme which ends the cellular respiratory chain, and is located in the mitochondrial membrane. Quantitatively important is the difference between the absorption spectra of the oxidised and reduced forms of CtOx. The concentration of cytochrome oxidase in living tissue is usually at least an order of magnitude below that of hemoglobin (Sato et al., 1976); therefore, its contribution is often neglected.

In the last 20 years there was an enormous growth in the instrumentation and applications of NIRS. A separate section will be devoted to the instrumentation. We will compare two NIRS devices: the INVOS Brain Oximeter (IBO) of Somanetics and the HEG device of H. Toomim.

## 3. Oxygen utilization

The amount of oxygen in the arterial blood depends upon the inspired oxygen and the pulmonary gas exchange. It depends on the arterial blood gas partial pressures of oxygen ($PaO_2$) and carbon dioxide ($PaCO_2$). The units of partial pressures may cause some confusion. Three types of units are in use. One unit is kiloPascal (kPa) equivalent to 7.5006 mmHg (or Torr). It can be measured also in %, when 100% corresponds to atmospheric pressure of 760 mmHg (Torr), i.e 1% corresponds to 7.6 Torr (mmHg). The arterial hemoglobin saturation ($SaO2$) is measured in %. The normal value is about 95%. A typical oxygen carrying capacity of the blood is 19.4 ml of $O_2$ per dl of blood with 19.1 ml $O_2$/dl carried by hemoglobin and only 0.3 ml $O_2$/dl dissolved in plasma (Cope, 1991). It should be noted that the oxygen delivery to the tissues is by diffusion and the hemoglobin acts as a buffer to maintain plasma's oxygen which is extracted by the tissue.

A typical averaged value for adult cerebral blood flow (CBF) is 47.7 ml/100 ml/min (Frackowiak et al., 1980) corresponding to total oxygen delivery 9.25 ml O2/100 ml/min. (Cope, 1991). Typical oxygen consumption of the adult brain is 4.2 ml O2/100 ml/min (Frackowiak et al., 1980). CBF, cerebral blood volume (CBV) and cerebral oxygen extraction (COE) are significantly greater in grey matter compared to white matter in normal human adults (Lammertsma et al., 1983; cope, 1991). The CBF and the cerebral blood volume (CBV) of grey matter in normal human adults is approximately 2.5 times that of white matter, while the cerebral oxygen extractions (COE) are 0.37 and 0.41 for grey and white matter respectively (Lammertsma et al., 1983; cope, 1991). Only part of the arterial oxygen which arrives in the brain is absorbed and utilized. The fraction which is utilized, known as the oxygen extraction fraction (OEF), is defined as

$$OEF = (SaO_2 - SvO_2)/SaO_2,$$

where $SaO_2$ and $SvO_2$ are the arterial and venous oxygen saturations respectively.

According to Derdeyn et al (Derdeyn et al, 2002) the EOF, measured in their normal control subjects, was 0.41±0.03. Assuming $SaO_2$ equal to 0.95 the $SvO_2$ will be, using Eq. (6), equal to 0.56±0.03. In the brain tissue the absorption of the oxygenated and deoxygenated hemoglobin is mostly venous. Assuming a 75% venous contribution, the brain tissue regional oxygen saturation ($rSO_2$) in the frontal region will be about 66±3%, a mean value which is observed in experiments. With a decrease of CBF there is a bigger demand for oxygen and EOF will increase (Kissack et al, 2005). Accordingly, with an increase of CBF the OEF will decrease.

## 4. NIRS Instrumentation

Several types of NIRS equipment, based on different methods, are commercially available. They measure the concentrations of Hb, $HbO_2$ and the total hemoglobin tHb. If the redox state of CytOx is also taken into account, then measurements with 3 wavelengths has to be done. Instruments with 2 wavelengths do not evaluate the CytOx contribution.

Three types of instruments are in use according to the used method : continuous intensity, time resolved and intensity modulated. For details, see (Delpy and Cope, 1997). Most of the commercial instruments utilize continuous wave (CW) light. In combination with the modified Lambert-Beer law it allows to measure changes in Hb and $HbO_2$. In a biological tissue, quantification of the NIRS signal is difficult. Different methods have been proposed to improve the resolution. One of them is the spatially resolved spectroscopy (SRS), which uses CW light and a multi-distance approach. With this method the $rSO_2$ (the absolute ratio of $HbO_2$ to the total Hb content-tHb), can be evaluated (Suzuki et al., 1999).

A further distinction among the instruments can be made. The simplest are the photometers, which use single-distance and CW, light, usually with one sensor (channel).

The oximeters are more sophisticated. They use multi-distance (SRS) techniques with CW and usually two sensors (channels). For details, see (Ferrari et al., 2004; Delpy and Cope, 1997; Rolfe, 2000). Recently several groups have begun to use multi-channel CW imaging systems generating images of a larger area of the subject's head with high temporal resolution up to 10 Hz (Ferrari et al., 2004; Miura et al., 2001; Obrig and Villringer, 2003; Quaresima el al., 2001a. 2002a).

In the following we will compare two instruments: the INVOS Cerebral Oximeter of Somanetics (www.somanetics.com; Thavasothy et al., 2002), and the hemoencelograph (HEG) (Toomim and Marsh, 1999; Toomim et al., 2004).

### 4.1 The INVOS oximeter

The Somanetics INVOS Cerebral Oximeter (ICO) system measures regional hemoglobin oxygen saturation ($rSO_2$) of the brain in the area underlying the sensor and uses two wavelengths, 730 and 810 nm. The sensor, ( "SomaSensor"), is applied to the forehead with an integrated medical-grade adhesive. Two sensors can be placed in the forehead near Fp1 and Fp2. The spatially resolved spectroscopy (SRS) method is applied by using in the sensor two source-detector distances: a "near" (shallow), 3 cm from the source and a "far" (deep), 4 cm from the source. Both sample almost equally the shallow layers in the tissue volumes directly under the light sources and detectors in the sensor, but the distant "far" penetrates deeper into the brain. Using the SRS method, subtraction of the near signal from the far should leave a signal originating predominantly in the brain cortex. The measurement takes place in real time, providing an immediate indication of a change in the critical balance of oxygen delivery and oxygen consumption.

According to the producers: "Using the model at a 4 cm source-detector spacing and no signal subtraction, the overlying tissue and skull contribute, on average, about 45 percent of the signal while 55% is cerebral in origin. Subtracting the data from the 3 cm spacing (as the Oximeter does) reduces this extracerebral contribution to less than 15 percent. While the potential exists to develop an instrument that will reduce the extracerebral contribution to zero, subject-dependent variations in anatomy and physiology will likely cause variations of ±10%. While the extracerebral contribution is not zero, the noninvasive Somanetics INVOS Cerebral Oximeter provides a "predominately cerebral" measurement where over 85 percent of the signal, on average, is exclusively from the brain" (www.somanetics.com).

The INVOS Cerebral Oximeter is an important tool in surgery rooms, saving lives and expenses. The producers explain: " Declining cerebral oximeter values occur frequently in cardiac surgery and reflect the changing haemodynamic profile of the balance between brain oxygen delivery and consumption. Since low rSO2 values correlate with adverse neurological and other outcomes, continuous assessment is a valuable patient management tool. Declining or low cerebral oximeter values are corrected with simple interventions".

**4.2 The HEG**

The hemoencephalograph (HEG) is a single-distance CW spectrophotometer, which uses NIR light with two wavelengths, 660 and 850 nm. The light source consist of closely spaced emitting diodes (LED optodes). The source and an optode light receiver are mounted on a headband. The distance between the source and receiver is 3 cm.

The HEG measures the ratio of the intensity of the 660 nm light to the intensity of the 850 nm light.

The HEG is not intended to measure $rSO_2$. Nevertheless it is an important tool in the biofeedback research.

Hershel Toomim, the inventor of HEG has noticed that he can influence the outcome by looking at the results. Since then many people were able to increase the HEG readings via such a biofeedback.

The HEG became an important tool for training local brain oxygenation. The HEG is a very sensitive device. The distance between the source and receiver is the same as the distance of the shallow detector of the Somanetics INVOS Cerebral Oximeter. Therefore the INVOS Oximeter covers larger brain tissue and is more stable and less influenced by biofeedback.

What is the HEG measuring? According to the producers: "The HEG ratio is the basis of blood flow training. A normalized basis for HEG was established using measurements at Fp1 of 154 adult attendees at professional society meetings. A normalized reference value of 100 (SD=20) was thus established and served to calibrate all further spectrophotometers". The calibration was achieved by multiplying the intensities ratio by 200. We have shown (Gersten et. al. 2006) that one can relate the readings of the HEG to $rSO_2$ and even calibrate it separately for each individual.

## 5. Experimental detection of oxygen waveforms

We started this research in Israel in the Pulmonary Unit of the Soroka University Hospital in Beer-Sheva. The study protocol was approved by the Helsinki (Ethics) Committee of the Soroka University Hospital. The investigation conforms with the principles outlined in the Declaration of Helsinki. The nature of the study was explained, and all subjects gave written consent to participate.

At the beginning our research was based on capnometer measurements of end tidal $CO_2$ ($EtCO_2$). The capnometer measures the $CO_2$ concentration of the expired air. During the inspiration or breath holding the capnometer indications are zero. The capnometer enable us to follow the breathing periodicity.

We continued the research in the USA using the INVOS Cerebral Oximeter model 5100B of Somanetics Corporation and a capnometer of Better Physiology, Ltd. Both devices are noninvasive. The INVOS Cerebral Oximeter is based on most recent technological developments of near infrared spectroscopy (NIRS). With this device data are collected of regional oxygen saturation ($rSO_2$) near the forehead with two optical sensors (for more details see www.somanetics.com). In addition to $rSO_2$ one can determine the Blood Volume Index (BVI), which is an indicator of blood changes in the brain. This is a relative quantity, which could not be normalized with our oximeter.

The fastest recording rate of the 5100B oximeter is every 12 seconds from both left and right sensors. This time is much longer then the average period of about 4 seconds of normal respiration. In order to detect oxygenation periodicity we had to study respiration periods of about 36 seconds or larger (i.e. 3 data points or more for each breathing period). This is still a rather small amount of data points per respiration period. We have compensated for this small number by using a cubic spline interpolation of the data points, adding new interpolation points through this method. The cubic spline interpolation is a very effective method of smooth interpolation.

We found six people well acquainted with yoga pranayama, who could easily perform breathing exercises with periods around 36 seconds. All of them performed the following routine which lasted for 15 minutes.

They were asked to breathe in the following way: to inhale for 4 units of time (UOT), to hold the breath for 16 UOT and to exhale for 8 UOT, this we denote as the 4:16:8 (pranayama) routine.

The unit of time (UOT) is about 1 second. The yoga practitioners develop an internal feeling of UOT which they employ in their practices. They learn to feel their pulse or they learn to count in a constant pace. Often they practice with eyes closed. In order not to distract or induce additional stress we preferred not to supply an external uniform UOT. The primary concern for this research was to have a constant periodicity and in this case the practitioners have succeeded to maintain it. The data were analyzed with spectral analysis which took into account non-stationary developments, which were subtracted from $rSO_2$ and BVI data. Sharp picks corresponding to the breathing periodicity were found in the spectral analysis of $EtCO_2$ (the amount of $CO_2$ during expiration).

The motivation for this exercise was to see the relation of $rSO_2$ (oxygenation) to the increasing amount of $PaCO_2$. Actually instead of $PaCO_2$ the end tidal CO2 (EtCO2) was measured (the maximal $CO_2$ at exhalation), a quantity which approximate well the $PaCO_2$. The periodicity of rSO2 was studied during the 4:16:8 respiration period (of approximately 32 UOT).

The $rSO_2$ was measured with the aid of two sensors placed on the forehead, detecting oxygenation from the left hemisphere (frontal part in about 3 cm depth) and the right hemisphere (frontal part in about 3 cm depth). At the same time data for evaluation of BVI were collected. In Fig. 2 the results of simultaneous measurements of $EtCO_2$, $rSO_2$ and differences in BVI are presented for the 6 subjects.

In order to check more precisely the periodic behavior, spectral analysis was performed on the $EtCO_2$ data and the interpolated $rSO_2$ and BVI data. The $rSO_2$ and BVI data have a non-stationary component. The spectral analysis was performed on the raw $EtCO_2$ data and on $rSO_2$ and BVI interpolated data from which the non-stationary components were subtracted. The results are shown in Fig. 2a., Fig. 2b, Fig. 2c, Fig. 2d, Fig. 2e, Fig. 2f and Table 1. There is a clear overlapping between the periodicity of $EtCO_2$ and the periodicities of $rSO_2$ and BVI.

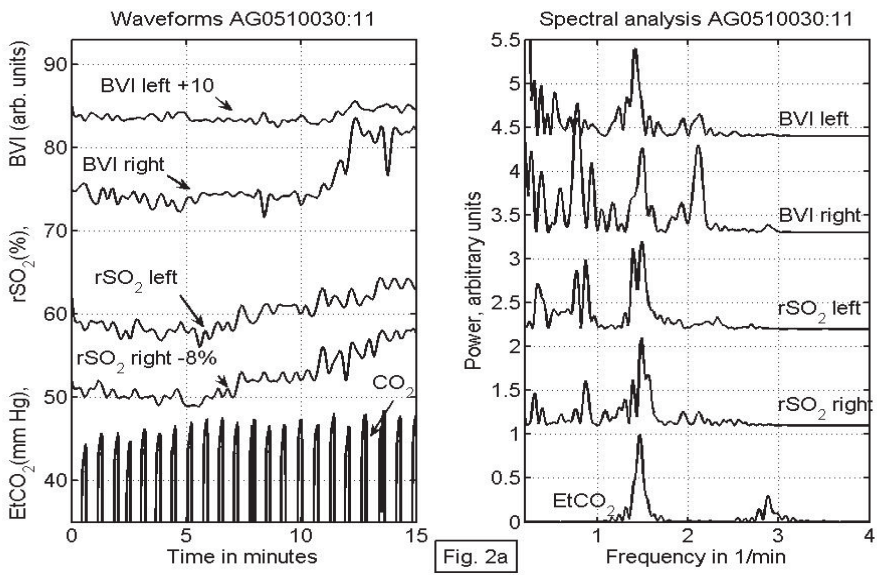

Fig. 2a. On the left hand side are given from the bottom to top: the readings of the capnometer in mm Hg, $rSO_2$ from the right sensor (subtracted with 8%), $rSO_2$ from the left sensor, the difference of BVI from the right sensor, and the difference of BVI from the left sensor (increased by 10). On the right hand side the corresponding spectral analyses of the waveforms are given.

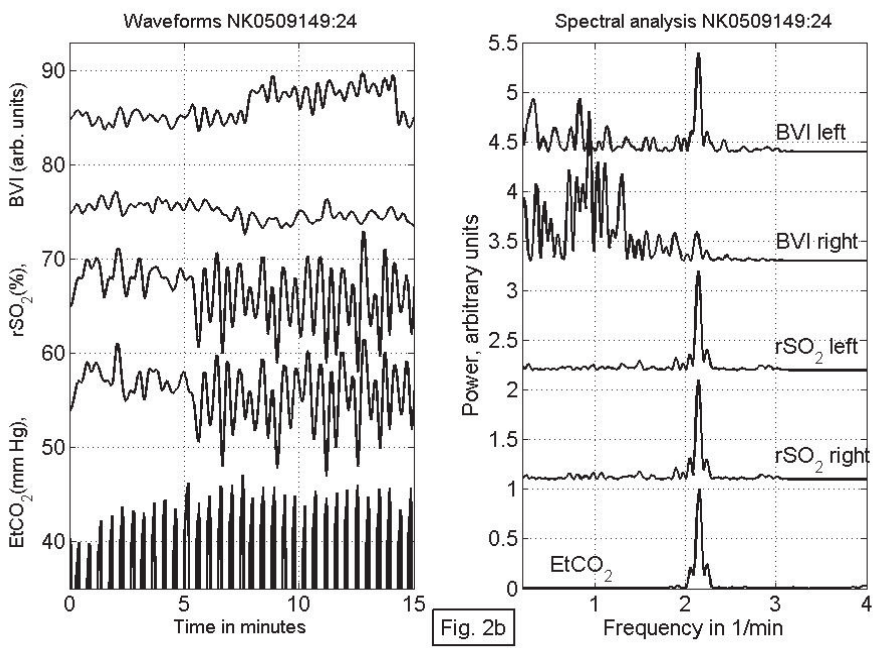

Fig. 2b. As Fig.2a.

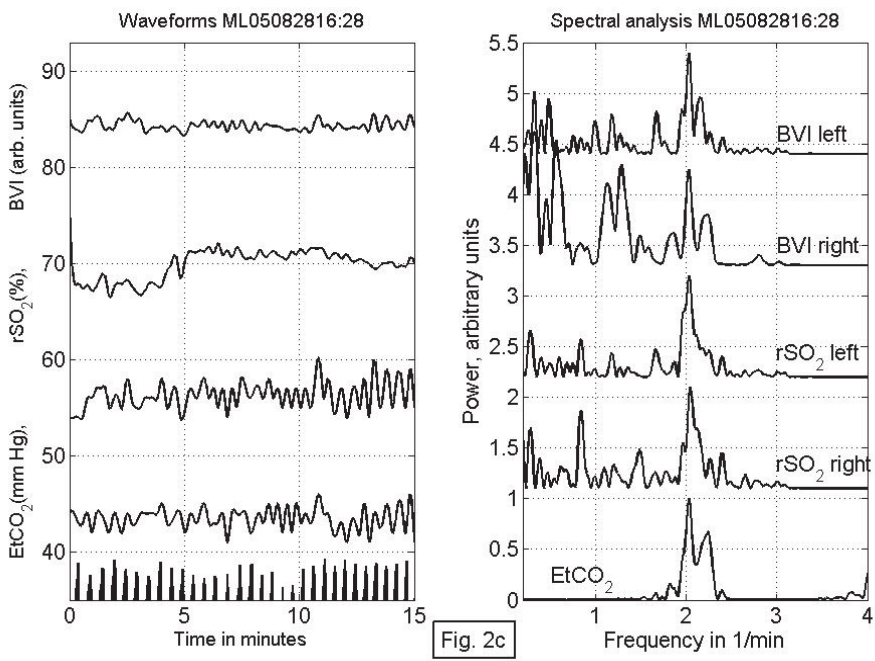

Fig. 2c. As Fig.2a.

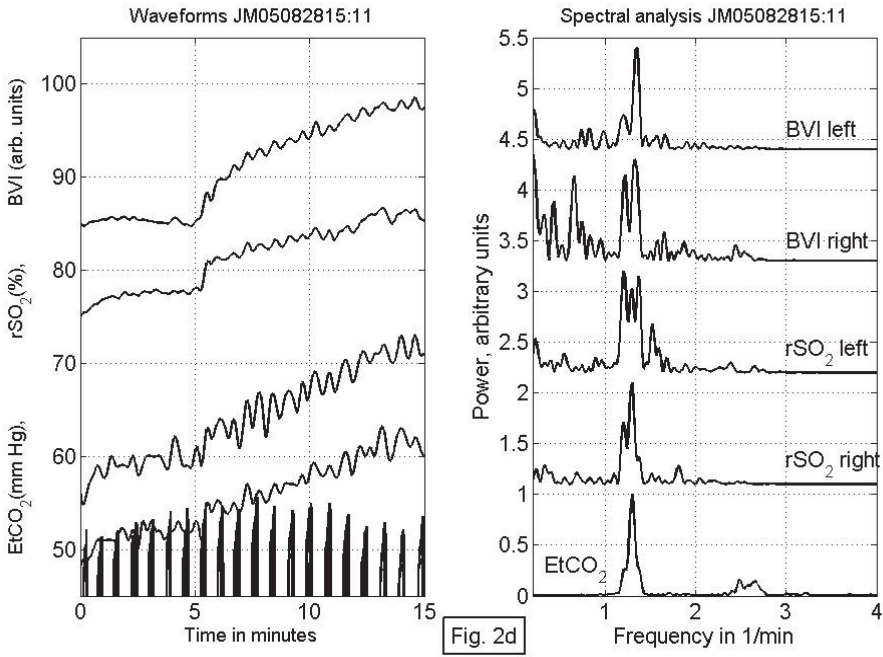

Fig. 2d. As Fig.2a.

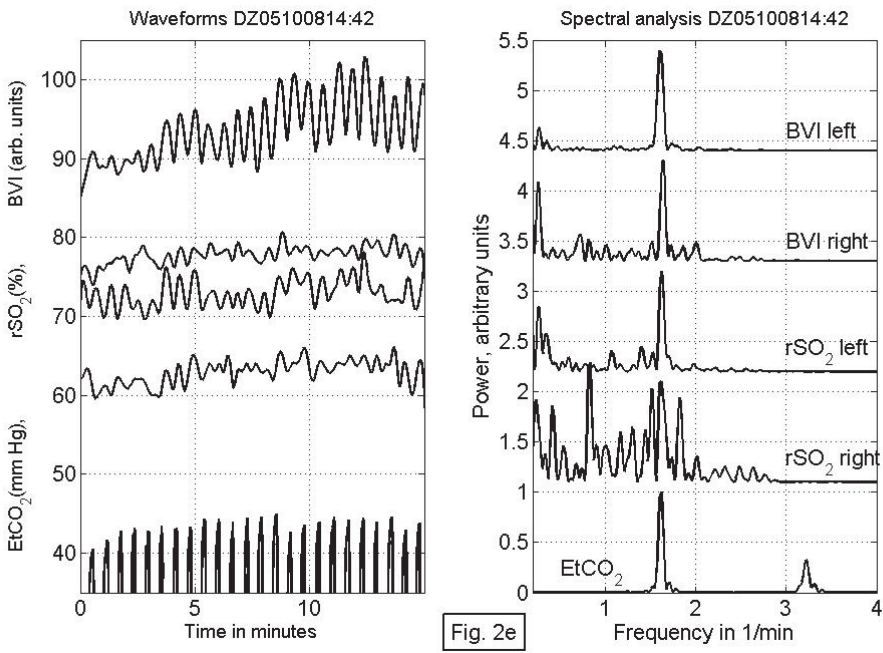

Fig. 2e. As Fig.2a.

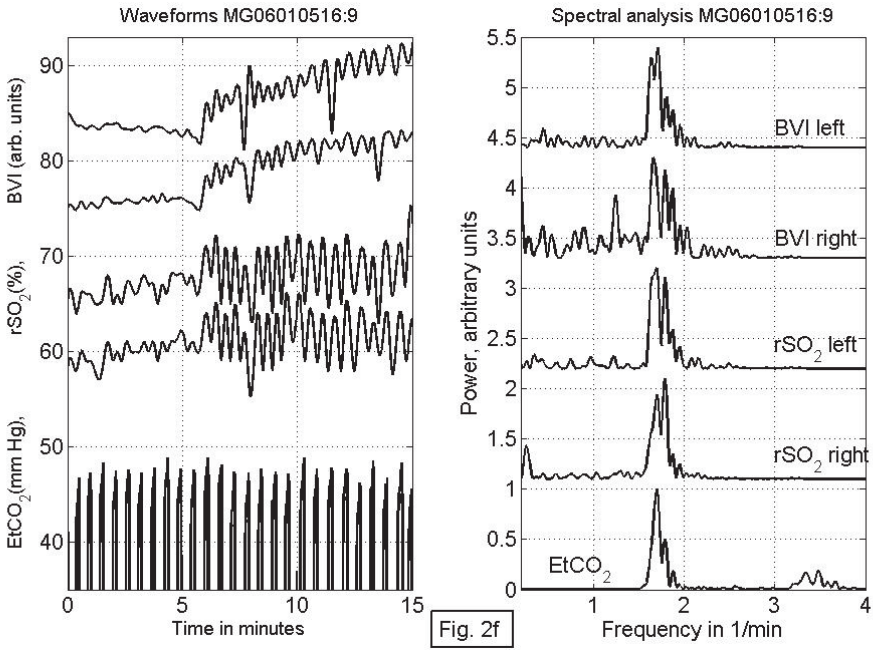

Fig. 2f. As Fig.2a.

|     | PaCO$_2$ | rSO$_2$ right | rSO$_2$ left | BVI right | BVI left |
|-----|----------|---------------|--------------|-----------|----------|
| **AG** | 1.47 (1.42-1.51) | 1.49 (1.45-1.53) | 1.48 (1.45-1.54) | 1.49 (1.42-1.53) | 1.42 (1.38-1.47) |
| **NK** | 2.15 (2.12-2.18) | 2.14 (2.12-2.17) | 2.14 (2.12-2.17) | ------------------- | 2.14 (2.11-2.17) |
| **ML** | 2.03 (1.98-2.07) | 2.05 (2.01-2.15) | 2.03 (1.95-2.08) | 2.03 (1.99-2.07) | 2.03 (2.00-2.06) |
| **JM** | 1.30 (1.26-1.34) | 1.30 (1.26-1.33) | 1.30 (1.17-1.40) | 1.33 (1.29-1.37) | 1.35 (1.31-1.38) |
| **DZ** | 1.62 (1.59-1.65) | 1.62 (1.59-1.67) | 1.62 (1.60-1.65) | 1.63 (1.61-1.67) | 1.61 (1.57-1.65) |
| **DG** | 1.70 (1.65-1.73) | 1.79 (1.65-1.82) | 1.70 (1.62-1.82) | 1.66 (1.62-1.89) | 1.71 (1.61-1.74) |

Table 1. The position of the dominant frequencies (in units of 1/minute) of Fig. 2 in the spectral analysis, in parenthesis the extension of the half width is given.

## 6. Simple exercises

Can simple exercises be devised to increase cerebral blood flow (CBF) and/or cerebral oxygenation? We investigated exactly that question by using three different techniques, namely: a simple breathing procedure, solving an arithmetic problem and biofeedback.

Elsewhere (Gersten , 2011) we have analyzed the influence of arterial partial pressure of $CO_2$ (PaCO$_2$) on CBF and found that it may dramatically change the CBF. The changes involve the blood flow of the whole brain. It is a global effect. These results were used in another investigation (Gersten et al., 2011) in which yoga practitioners were increasing their PaCO$_2$ through periodic yoga (pranayama) breathing techniques.

We will demonstrate that significant increase of $PaCO_2$ (and of total CBF) can be achieved with untrained people using very simple breathing procedures. The reason for that is the dependence of $PaCO_2$ on ventilation (West, 1992)

$$PaCO_2 = K \frac{\dot{V}_{CO2}}{\dot{V}_A}, \quad K = 863 \, mm \, Hg \tag{1}$$

where $\dot{V}_{CO2}$ is the $CO_2$ production (dependent on the metabolism) and $\dot{V}_A$ is the alveolar ventilation. This means that the $PaCO_2$ is inversely proportional to ventilation. Therefore it is possible to control the $PaCO_2$ by either breathing slowly (and not increasing the tidal volume substantially) or by holding the breath. Untrained people increase their tidal volume while breathing more slowly, but the overall effect is usually a slight increase in $PaCO_2$. People trained in breathing exercises may increase their $PaCO_2$ considerably by learning to control their tidal volume. For very small ventilation a correction is needed to the $PaCO_2$ formula (Riggs, 1970),

$$PaCO_2 = K \frac{\dot{V}_{CO2}}{\dot{V}_A - \dot{V}_D}, \quad \dot{V}_D = 2.07 \, l/min, \tag{2}$$

where $\dot{V}_D$ is the contribution of the dead space.

It is well known that concentrating on a mental problem changes brain's oxygenation locally (Chance et al, 1993). This finding was also applied further to solving a simple arithmetic problem on local brain oxygenation near the Fp1 area.

Hershel Toomim (Toomim et al., 2004) developed the device called hemoencephalograph (HEG), whose readings is related to regional cerebral oxygenation (Gersten et. al, 2007c). The device has many advantageous features which allowed us to use it in our experiment. Toomim has observed that he can influence the results by looking at the HEG display, which is essentially a biofeedback technique used by us as well. As a result of Toomim findings many biofeedback experiments were conducted with the HEG, confirming the effect. The readings of the HEG are very sensitive to changes in the range of normal oxygenation of the brain. This is not the case with INVOS brain oximeters used in operation rooms whose main aim is to detect abnormally low oxygenation states. For that reason we preferred to use the HEG to detect biofeedback effects even though it is much simpler and less sophisticated compared with INVOS cerebral oximeters (Gersten et al., 2009).

The readings of the HEG are normalized to 100 (SD=20), the average on 154 adult attendants at professional meetings (Toomim et al, 2004). We have compared (Gersten et al., 2009) the readings of HEG with the regional saturation of oxygen ($rSO_2$) readings of the INVOS cerebral oximeter of Somanetics. This allowed us to make estimates of the ratios of $rSO_2$ using the HEG. We found,

$$x_1/x_2 = \ln(y_1/32.08)/\ln(y_2/32.08), \quad x \equiv rSO_2, \quad y \equiv \text{HEG readings}, \tag{3}$$

where ln is the natural logarithm, but the ratio of logarithms does not depend on logarithm's basis.

Measurements were taken using HEG and a capnometer (a device measuring end tidal $CO_2$) simultaneously. End tidal $CO_2$ is closely related to $PaCO_2$. Eighteen subjects participated in the experiment in which HEG and $CO_2$ data were recorded for 5 intervals of baseline, simple breathing exercises, simple arithmetic tasks and biofeedback. The results show that almost all participants could increase their brain oxygenation or CBF, but in each case it was strongly dependent on one of the three methods used. We can conclude that it is possible to substantially increase local oxygenation or global CBF using one of the three methods described above, but the preferred method is highly individual. The protocol of this research was approved by the IRB of Hunter College of the City University of New York.

## 6.1 Methods and materials

It is well known that breathing patterns affect the $CO_2$ levels in the arteries (Fried and Grimaldi, 1993), which in turn can affect cerebral (brain's) blood circulation and oxygenation. Mental work and biofeedback may affect both local as well as global oxygen levels in the brain.

The influence of breathing exercises, problem solving and biofeedback on brain oxygen and CO2 arterial levels were considered in an experiment outlined below. The experiment dealt with 3 topics

1. The physiological effects of mild breathing exercises on increasing $CO_2$ and oxygen levels in the brain.
2. The physiological effects of problem solving (a particular case of mental performance) on the $CO_2$ and oxygen levels in the brain.
3. The physiological effects of biofeedback on the $CO_2$ and oxygen levels in the brain.

An experiment dealing with the second and third topic gives qualitative information about how much the oxygen levels will rise during problem solving and biofeedback, while an experiment dealing with the first topic will give the same information, but this time, from breathing exercises. It is important to note that in the first topic global CBF is concerned, while in the second and third topic the local brain oxygenation at the Fp1 area.

The participants were connected to the two devices needed for the experiment: the capnometer that measured end tidal $CO_2$ and the Cerebral Oximeter. The connection used was made via a sensor placed on the forehead at Fp1. The $CO_2$ levels were estimated using a capnometer produced by Better Physiology LTD, which measures end tidal $CO_2$ ($EtCO_2$) of the exhaled air ($EtCO_2$ is highly correlated with the $PaCO_2$ levels of the arteries). All participants received their own new nasal insert which were sterilized before each use and connected to the capnometer. The data were detected via USB output cable connected to the computer and stored for subsequent review.

The oxygen levels were estimated using two devices: the HEG which was calibrated to the INVOS Cerebral Oximeter produced by Somanetics Corp. (see www.somanetics.com) and based on advanced near infrared spectroscopy (NIRS) technology. A sensor was attached to the forehead measuring the oxygenation in a depth of about one inch inside the brain. The devices that were used were non-invasive and FDA approved, fully automated and did not require special precautions. The data were stored on a computer.

The data of the capnometer and oximeters were combined together and analyzed using Matlab subprograms. The participants were asked to do paced breathing exercises as instructed by the experimenters. Before the 3 experiments baseline data were taken for 5 minutes using the INVOS oximeter, and another 5 minutes using the HEG and capnometer. The participants were prevented from seeing the screens of the devices in order to avoid biofeedback.

In the first experiment (lasting 5 minutes) participants were asked to walk slowly, breathe in for 3 steps, hold their breath during the next 3 steps, exhale during the next 3 steps, and hold their breath for the next 3 steps after exhaling. We made sure that the participants understood these instructions. The participants also did not see the screens of the devices in order to avoid biofeedback.

In the second experiment (lasting 5 minutes) participants were given an arithmetical problem to solve while being attached to the HEG and capnometer. The theoretical basis for this experiment is that more oxygen is needed while solving problems. A simple arithmetical problem of subtracting the number 7 continuously, starting from 1200 (1193,1186,...) was used. The participants were again prevented to see the screens of the devices in order to avoid biofeedback.

In the third experiment (lasting 5 minutes) participants were asked to look at the HEG display trying to raise the curve by mental feedback. This time they were allowed to look at the display.

### 6.2 Participants

The participants were 18 participants from the introductory course to psychology (PSY 100) in Hunter College of the City University of New York. All participants had to sign an informed consent. At least two experimenters were present during each experiment. The confidentiality of the participants was protected.

### 6.3 Results

All experimental results of all participants can be found in (Gersten et al., 2011a).

To better illustrate the results a few examples of HEG and $CO_2$ data will be included. In Fig. 3, subplot $HC3_1$, the baseline data of participant No. 3 are displayed. The HEG baseline was not constant during the 5 minutes of data taking.

In the subplot $HC3_2$ the result of the breathing exercise are displayed. Even though this was a first trial, the $CO_2$ pattern seems to be quite periodic. The $CO_2$ pattern has a periodicity of about 15 secs per period, while normal breathing has a periodicity of about 4 secs per period. The prolongation of the respiratory period should lead to an accumulation of arterial $CO_2$ and an increase of global CBF, provided there is no greater increase in the tidal volume. In this case there was only small increase in arterial $CO_2$ but a significant increase in oxygenation (HEG). The increase of arterial $CO_2$ depends on the control over the tidal volume. Individuals trained in this breathing exercise, can easily increase their arterial $CO_2$ by about 20-30%.

Solving the arithmetic problem (see Fig. 3, subplot $HC3_3$) led to an increase of the HEG readings (oxygenation), but we notice that the respiration was speeded up probably due to increased tension. After all, to subtract 7 and calculate and evaluate the result in the mind is not a very pleasant enterprise.

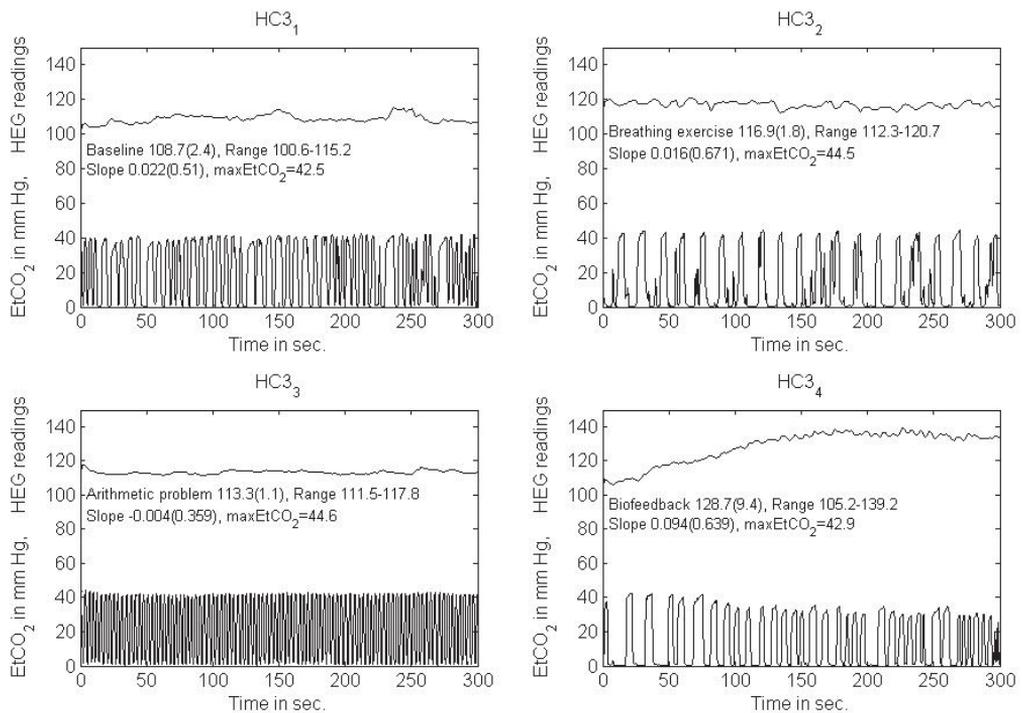

Fig. 3. The HEG readings and the capnometer displays (EtCO$_2$) of participant No. 3 are shown for the 4 cases: baseline (subplot HC3$_1$), breathing exercise (subplot HC3$_2$), arithmetic problem (subplot HC3$_3$) and biofeedback (subplot HC3$_4$). The upper curves are the HEG readings, the lower curves the CO$_2$ values (at exhale). The numbers with numbers in parenthesis are the mean values and standard deviations respectively.

Subplot HC3$_4$ of Fig. 3 is very interesting. In this case the subject was looking at the display of the HEG line trying mentally to raise it up. The performance (without previous training) is very impressive. Starting from baseline values the HEG readings were climbing up for about 2.5 minutes to values about 30% higher (a local increase of oxygenation at the Fp1 area). This case teaches us that in the evaluation of the results we must not only consider average values but also the maximal values which are an indicative of the possible potential. The respiration pattern indicates a slow but a very deep breathing (hyperventilation). The arterial CO$_2$ decreased for more than 20% which should lead to a significant decrease of global CBF. The biofeedback was very successful in spite of the hyperventilation.

While analyzing the data we must take into account that the participants were performing their tasks for the first time. Most of the tasks were performed relatively well. Most participants were able to increase their mean HEG readings in at least one task. When averaging all participant's data there were no significant changes were found, indicating that no one method was preferable.

More informative, are the ratios of the maxima to the mean of the baseline. The maxima indicate the potential of the exercises. These maximal values should be easily reached with practice. There was a maximal increase of 30% during breathing exercises, 32% during solving the arithmetic problem and 28% during the biofeedback.

Results with $rSO_2$ ratios determined according to Eq. (3) are quite similar to the HEG ratios, indicating that the HEG ratios are quite reliable in estimating the $rSO_2$ changes.

Of the 18 participants 14 were able to increase the HEG readings by at least 10% during one of the exercises (5 during the breathing exercise, 9 while solving the arithmetic problem, 8 during the biofeedback). Of the 18 participants 7 were able to increase the HEG readings by at least 18% during one of the exercises (3 during the breathing exercise, 6 while solving the arithmetic problem, 5 during the biofeedback).

The breathing exercise was the most difficult for the participants. It took them an average of 6 minutes to fully understand the instructions. The exercise required some discipline and experience. Most of the participants performed it relatively well.

Fig. 4 shows that participant No. 9 has performed the breathing exercises relatively well (subplot $HC9_2$). His $CO_2$ pattern was periodic and amplitude stable. The pattern was not completely smooth. This is understandable, since it was his first attempt to perform the exercise. In the same subplot, the corresponding HEG curve is very interesting. The HEG waveform has the same period as the $CO_2$ pattern.

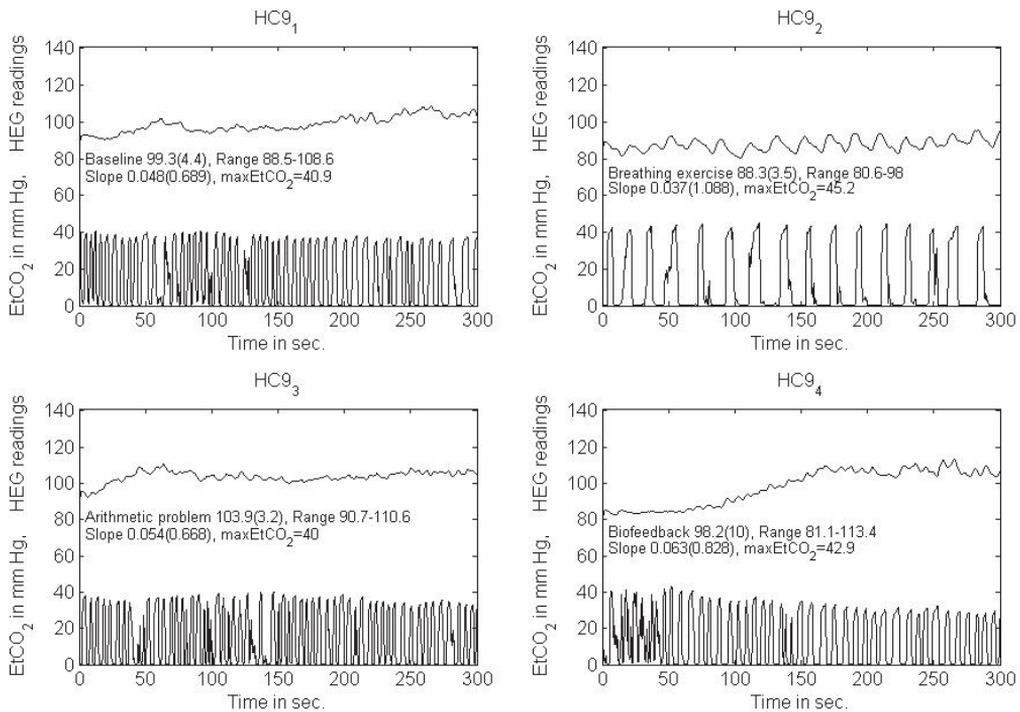

Fig. 4. The HEG readings and the capnometer displays ($EtCO_2$) are shown for the 4 cases: baseline (subplot $HC9_1$), breathing exercise (subplot $HC9_2$), arithmetic problem (subplot $HC9_3$) and biofeedback (subplot $HC9_4$). The upper curves are the HEG readings, the lower curves the $CO_2$ values (at exhale).

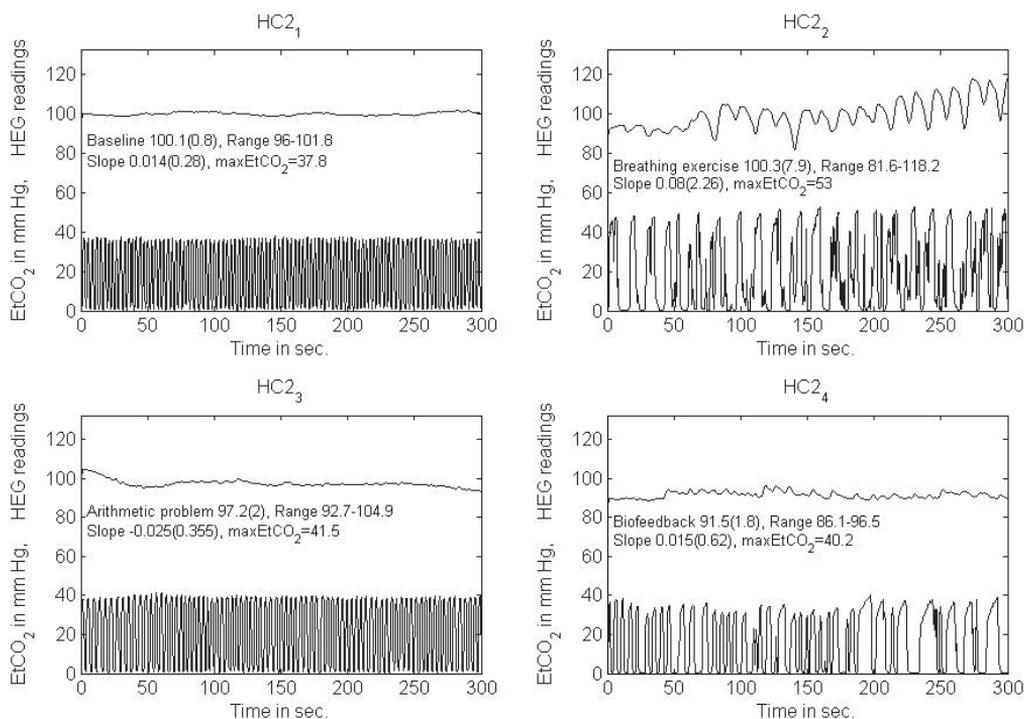

Fig. 5. The HEG readings and the capnometer displays (EtCO$_2$) are shown for the 4 cases: baseline (subplot HC2$_1$), breathing exercise (subplot HC2$_2$), arithmetic problem (subplot HC2$_3$) and biofeedback (subplot HC2$_4$). The upper curves are the HEG readings, the lower curves the CO$_2$ values (at exhale).

In Fig. 5 participant No. 2 had difficulty performing the breathing exercise (the CO$_2$ pattern in subplot HC2$_2$). Although the 3 step pattern was kept correctly, the participant was inhaling during some of the breath holding periods. When done correctly the 3 step breathing cycle should last for about 15 seconds. As the participant was breathing in between, the average breath length was only 7.2 seconds. Interestingly the HEG waveform of subplot HC2$_2$ has a period of about 15 seconds irrespective of the breathing in between the 3 step pattern.

The performance of the breathing exercises, when well performed, the period should be longer than 10 seconds. Twelve of 18 participants have performed the breathing exercise well.

Interestingly the HEG waveform has the same periodicity as the CO$_2$ pattern. This coincidence is well presented in Fig. 6, where the correlation between the power spectra of the EtCO$_2$ periodic pattern and the corresponding HEG periodic pattern is depicted by their multiplication. The power spectra are normalized to unity. Maximal correlation is achieved when the multiplication is equal to one.

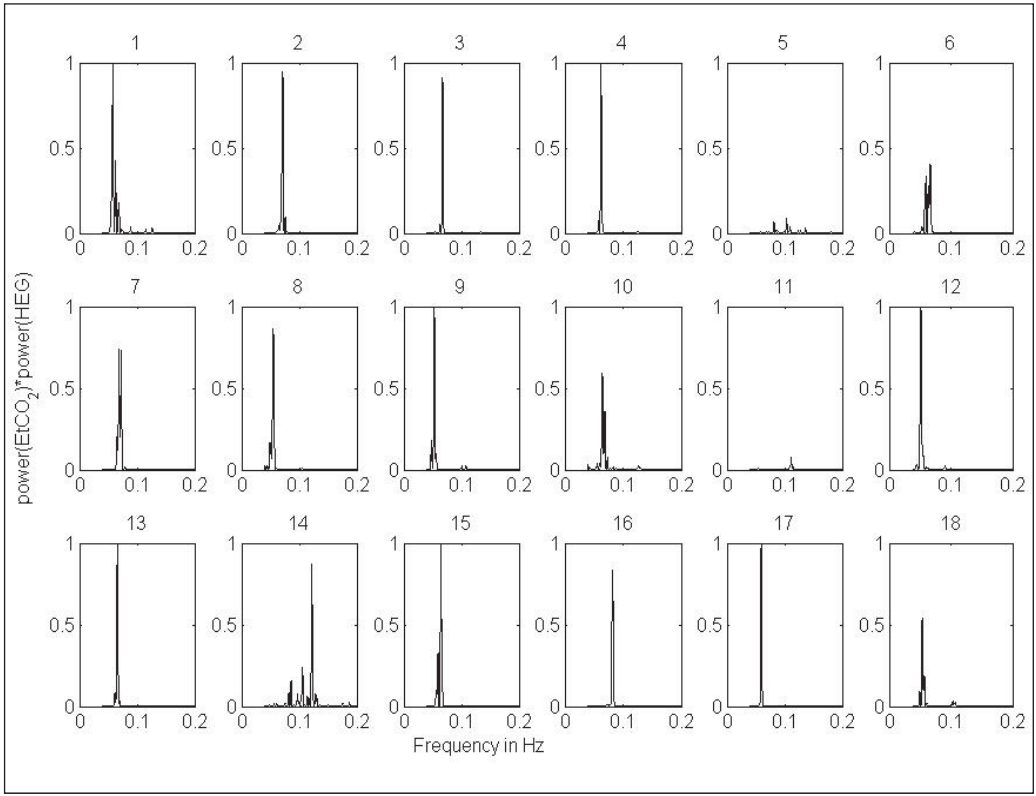

Fig. 6. The correlation between the power spectra of the EtCO$_2$ periodic pattern and the corresponding HEG periodic pattern is depicted by their multiplication. The power spectra are normalized to unity. Maximal correlation is achieved when the multiplication is equal to one.

## 7. Conclusions

The near infrared spectroscopy (NIRS) is a powerful non-invasive method that probes brain oxygenation. The subject is of great interest to researchers and the medical profession, resulting in many clinical, neuroscience and physiology of exercise studies and publications. NIRS is a noninvasive and easy to handle method, and will provide a new direction for functional mapping.

We have verified that breathing may affect to great extent the brain physiology. With breathing exercises the arterial CO$_2$ can be increased, which in turn increases CBF and brain oxygenation.

The most important finding of our experiments is the periodic correlation between respiration, oxygenation and blood volume changes. The results clearly show a periodic change of cerebral oxygenation with the same period as the breathing exercises, indicating that with each breath the brain oxygenation was periodically changing. Similar periodic changes in blood volume indicate that the brain pulsates with a frequency of respiration.

Present results were achieved by using very slow breathing patterns with the INVOS Cerebral Oximeter (ICO). If the present ICO devices will be modified to allow sampling of $rSO_2$ at frequencies higher than the frequency of normal respiration and higher than the heart rate, it will be possible to observe new types of brain waveforms. These waveforms may have new information about brain oxygenation, cognitive function, brain pulsation and brain motion. Under these circumstances it will be possible to understand much better the correlations between respiration and brain physiology. The sampling of $rSO_2$ should go up to, or preferably above 4 Hz. The accuracy of the reading device also should be changed from 2 significant digits to 3 digits. We believe that these changes will enable new explorations and new insights on the influence of respiration on brain's physiology. The HEG, which does not measure directly the $rSO_2$ can serve this purpose by using Eq. (3), which determines the ratios of $rSO_2$. The difference between the ICO and the HEG is that the ICO penetrates dipper into the brain, while the HEG penetrates only the surface near the probe. The ICO results are more stable and are related to larger volumes in the brain.

Neurodegenerative diseases are characterized by low CBF, in our research we have found effective ways to increase CBF. We will continue our research in order to explore the possibilities of increased CBF and its influences on intellectual abilities and on fighting degenerative diseases. Devices displaying the oxygenation periodic waveforms should be developed for new diagnostic and research purposes.

Our method to obtain the above results is through the use of human subjects. This is a new avenue in approaching the study of CBF, brain oxygenation, improving the cognitive function and especially in view of the growing elderly population.

Our three methods used in simple execises can be used on the general population, are non-invasive, without the use of pharmaceuticals and have no side effects. They differ from each other in that the breathing affects mostly the global blood flow, arithmetic problem solving and biofeedback affects the regional blood flow (in our case the Fp1 region).

Both our theoretical and experimental work differs from other studies due the specific instrumentation and our experimental procedure. Most of the results came close to our expectation.

We concluded that breathing can be used effectively to control CBF by the ventilatory control of end tidal $CO_2$. This research may have implications for complementary diagnosis and treatment of conditions involving regional cerebral metabolism such as cerebral vascular ischemia, seizures disorders, stroke, Alzheimer's disease, and more. Following that thought could lead us to improved cognitive function through a higher supply of oxygen to specific regions of the brain.

We foresee future more detailed investigations to be made in the area of the effect of $CO_2$ on specific regions of the brain. This would be of great interest because a higher $CO_2$ supply results in a higher blood flow and thus to more oxygen and better overall brain function, specifically cognitive function.